# An Agent-Based Self-Protective Method to Secure Communication between UAVs in Unmanned Aerial Vehicle Networks

Reza Fotohi[1*], Eslam Nazemi[1], Fereidoon Shams Aliee[1]

[1]Faculty of Computer Science and Engineering, Shahid Beheshti University, G. C. Evin, Tehran, Iran

*Corresponding author: Reza Fotohi, R_fotohi@sbu.ac.ir; Fotohi.reza@gmail.com*

**Abstract**

UAVNs (unmanned aerial vehicle networks) may become vulnerable to threats and attacks due to their characteristic features such as highly dynamic network topology, open-air wireless environments, and high mobility. Since previous work has focused on classical and metaheuristic-based approaches, none of these approaches have a self-adaptive approach. In this paper, the challenges and weaknesses of previous methods are examined in the form of a table. Furthermore, we propose an agent-based self-protective method (ASP-UAVN) for UAVNs that is based on the Human Immune System (HIS). In ASP-UAS, the safest route from the source UAV to the destination UAV is chosen according to a self-protective system. In this method, a multi-agent system using an Artificial Immune System (AIS) is employed to detect the attacking UAV and choose the safest route. In the proposed ASP-UAVN, the route request packet (RREQ) is initially transmitted from the source UAV to the destination UAV to detect the existing routes. Then, once the route reply packet (RREP) is received, a self-protective method using agents and the knowledge base is employed to choose the safest route and detect the attacking UAVs. The proposed ASP-UAVN has been validated and evaluated in two ways: simulation and theoretical analysis. The results of simulation evaluation and theory analysis showed that the ASP-UAS increases the Packet Delivery Rate (PDR) by more than 17.4, 20.8, and 25.91%, and detection rate by more than 17.2, 23.1, and 29.3%, and decreases the Packet Loss Rate (PLR) by more than 14.4, 16.8, and 20.21%, the false-positive and false-negative rate by more than 16.5, 25.3, and 31.21% those of SUAS-HIS, SFA and BRUIDS methods, respectively.

## 1  Introduction

An unmanned aerial vehicle (UAV) is, in fact, an aircraft flying with no human pilot on board. Instead, an operator or the on-board computer systems control autonomously its flight either remotely. By developments in computing, device miniaturization and communication, other flying objects including quadcopters, gliders, and balloons could be also included in UAVs. Historically, military operations utilized in missions imposing high-risk levels to human pilots. However, more applications were recently found in civilian domains for UAVs. They involve rescue and search operations, inspection, and policing [1,2, and 3]. Figure 1 and 2 shows two examples of Civil applications. The setup involves multiple components and numerous links to communication. The task of each link is to transmit certain kinds of information and data. Generally, based on the kind of transmitted information, 3 various types of links should exist in these networks, i.e. radio communication, Satellite link, and U2U. The radio communication links transmit telemetry data, control audio, and video information. Furthermore, the task of satellite links is to carry GPS, meteorological, and weather information, along with the data transferred by the radio communication links. UAV applications in the field of Civil have been added to the paper in detail as the following. Because of the high movement of the UAVs, their simple deployment, floating capability, and their low maintenance cost, they are usable in many civil applications. Moreover, using UAVs in most civil applications like real-time monitoring, wireless coverage, remote sensing, movie generation, goods delivery, search and rescue, precise farming, security and monitoring, and aerial photography are growing rapidly. Also, it is predicted that the civil infrastructures have 45 billion dollars of the market value of using UAVs. Now, using small UAVs has been expanding rapidly including a wide range of general and commercial works such as Entertainment, shipping, delivery, low enforcement, wildlife monitoring, help to search and rescue operations, gathering news, an inspection of pipelines and other infrastructures, estates, taking photo, geology, help the accidents, and entertainments.

Further discussion is provided in the following sections in this regard to prove the applications of UAVs with innate time sensitivity, and to indicate the insistence of offering security in communication channels [2]. Nevertheless, despite the advantages of UAVs in different applications as a result of the circumstances where the activities are monitored by no pilot, they are potentially susceptible to lethal threats. This strengthens the emergence of designing reliable and secure UASs and overcoming the challenges to prevent destruction and damage to other systems and human lives [3].

Hence, UAVs become a fascinating target of lethal attacks, theft, and manipulation. Some attacks including Sinkhole (SH), Wormhole (WH), and Selective Forwarding (SF) improperly enter the system. When an attack affects the unmanned system, it is difficult to remove the threat and bring the system back online. It is worth to mention that the usual approaches to secure information, like intrusion detection or encryption [4] are insufficient to deal with such risks. For elaborating, the stated outlines do not take into account the actuator and sensor measurements compatibility factor with the control mechanism and physical procedure of the UAV, which are considerable for the protection outline. The malevolent UAV is strong against 3 lethal attacks (SH, WH, and SF) within the ASP-UAVN proposed design, hence, the intrusive operations are rapidly recognized and eliminated from the or top-secret data surveillance spying missions. Within the suggested schema, the critical standards of service quality are improved such as PLR, PDR, detection rates, false-negative, and false-positive rates.

This study is mainly focused as follows:

- Analyzing the UAV network to discover unknown attacks launched by external or internal attackers

- Designing an efficient intrusion and self-protective detection system utilizing an AIS for unknown and known attacks in UAV.

- Providing a set of descriptions of the most related routing protocols in the literature accompanied by their disadvantages. Moreover, we performed a comparative investigation for examining the deficiencies between our suggested scheme and the evaluated protocols.

- Performing a set of simulations to investigate the realistic impacts of UAVNs environments over our suggested protocol. The efficiency of ASP-UAVN was demonstrated by the attained results.

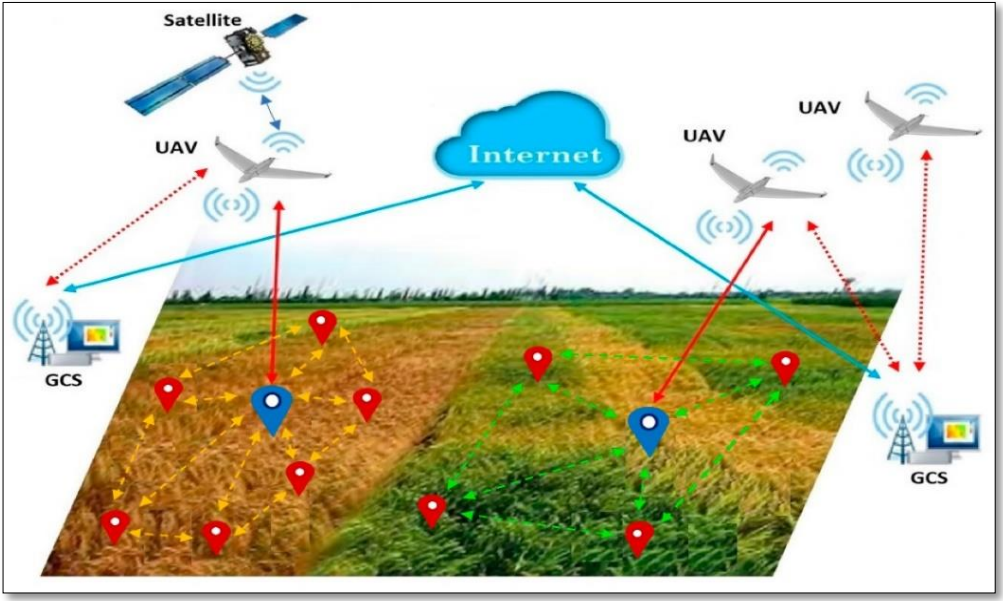

Figure 1: The first scenario of civil applications [5].

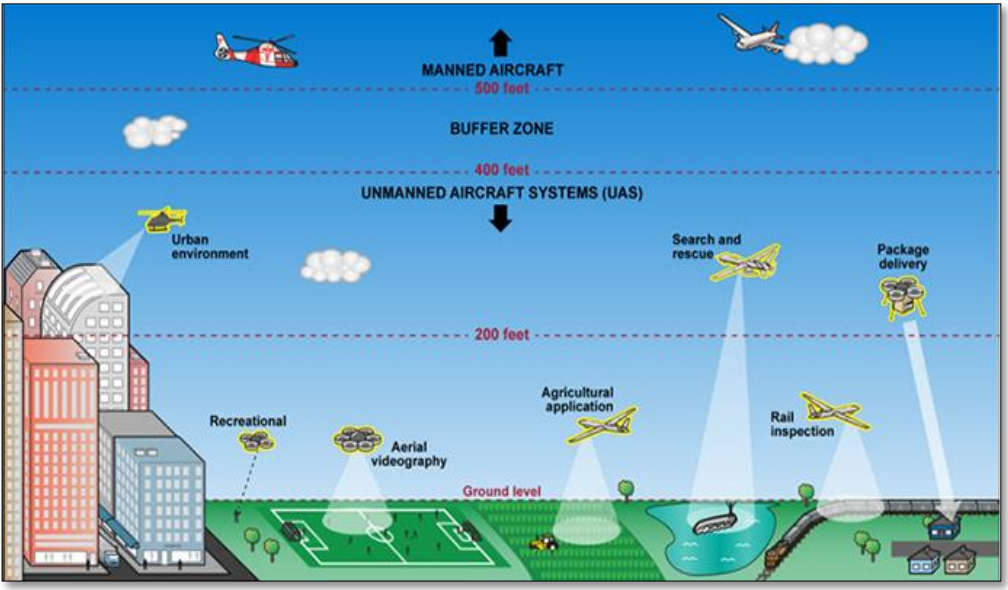

Figure 2: The second scenario of civil applications.

This paper is organized as follows: In Section 2, the security attacks and previous methods are explained in detail. Section 3, describes the human immune system (HIS) as an algorithm. The details of the proposed method are given in Section 4. In Section 5 of the paper, the theoretical security analysis of ASP-UAVN is described. In Section 6, the simulation results are discussed in the form of tables and figures. Finally, the conclusion and future work of this paper are done in section 7.

## 2 Lethal security threats and detection schemes

Here, the destructive attacks targeting UAVs, as well as previous methods used to defend against such destructive UAVs, are described in detail.

### 2.1 Lethal Security Threats

It was prone that UAVs are function degradation and cyber-security threats as active or passive since they depend on wireless channels for communicating. Figure 3 provides a list of main lethal security targeting UAVs. The following vulnerabilities are concerned with this study:

- *Wormhole Attack:* Or WH attacks are the main attack threatening the UAVs. In WH attacks, data packets are received by a hostile node at a definite location in UAV, and the packets are tunnelled to another hostile node at a distant point to regulate the packets to its adjacent nodes. It is possible to establish this tunnel using multiple techniques including a channel established out of band, a high-powered transmission, or an encapsulated packet. In these approaches, through tunnels, the packet transmitted is received rather directly or with fewer hop counts in comparison to ordinary packets that are conveyed via a multi-hop path. This method establishes an illusion whit two close tunnel endpoints [6]. Hence, the hostile nodes are made as decoys within the destination and source nodes that can accomplish subversions like packet manipulation and droppings.

- *Selective Forwarding Attack*: In this attack, a forged RREP is transmitted by an SF node while receiving an RREQ packet, appealing an unexpired and shorter route, even for missing the destination entry from the routing table. By reaching the created RREP packet the source node, a route is established via this malevolent middle node, to remove all legitimate RREP messages conveyed from destination nodes and another intermediate. Thus, the data traffic is successfully attracted by the BH node to that destination by misleading the source. Then, all the data packets are dropped by the SF node rather than forwarding the incoming messages. By forging a transmission route, the hop count is reset by the BH node to a very low value as well as the number of destination sequence to the quite high value to increment the reception opportunity at the source node. The SF attack can also launch from the source node through making fields-source sequence numbers in hop counts and RREQ packets, leading to harming the directing tables in middle nodes and the destination nodes [7].

- *Sink hole Attack:* One of the main attacks threatening the UAVNs is the attack known as the Sinkhole (SH) attack. In these attacks, a malicious node broadcasts illusive information regarding the routings to impose itself as a route towards specific nodes for the neighbouring nodes and thus, attract data traffic. The objective of this process is to draw all the traffic in the network towards the sinkhole node and as a result, alter the packets of data or silently drop them altogether. Sinkhole attacks can increase the network overhead, increase the consumption of energy and decrease the life time of the network , and ultimately annihilate the network [8].

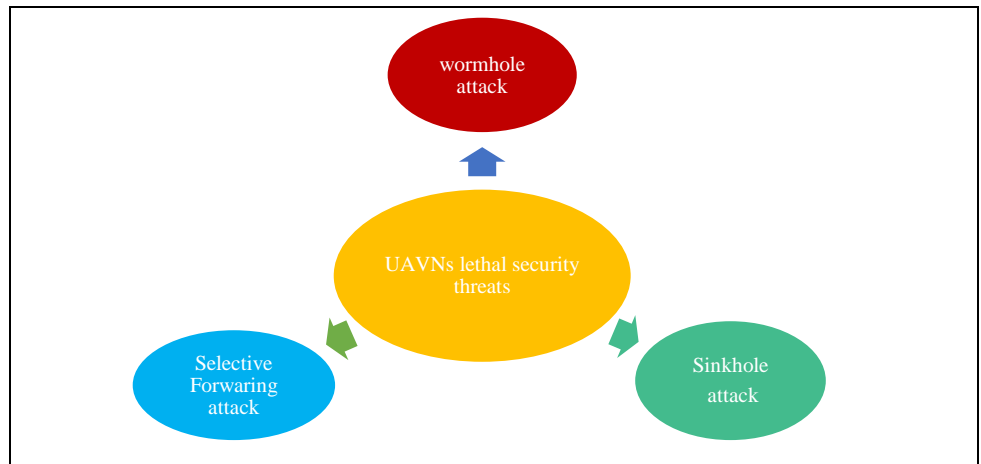

Figure 3: UAVNs lethal security threats

## 2.2 Detection schemes

Through different security measurements in different ways, lethal attacks were addressed and the UAV was protected against these attacks. It is not a new subject, and numerous studies were performed to provide various methods to state these attacks.

In [10], a security framework was proposed by the authors to offer protection against malicious performance targeting SFA communication systems in aircraft. According to the numerical outcomes presented in this work, the suggested security framework leads to prediction and detection rates with high accuracy in comparison with the intrusion detection approaches in the literature.

An adaptive IDS is suggested in [11], in terms of device specifications to find suspicious UAVs in cooperative operations with substantial operation continuity. the UAVs are audited by the suggested IDS system in a distributed system for determining their state whether normally functioning or under malicious attacks. In this study, the efficiency of the suggested rule-based UAV IDS (BRUIDS) performance is investigated on random, reckless, and opportunistic intrusive performances (usual cyber-attack behavioural techniques). The suggested technique is the base for the audition on behavioural rules to quickly investigate a UAV's survivability under malevolent attacks.

In [12], the authors suggest a way to improve drone security against wormhole, black hole, grayhole and fake information dissemination attacks so that communications can be conducted securely. This has two advantages: First, it has a high accuracy of detection and also a low false positive and negative rate. Second, it quickly detects and isolates attacks.

This study supports using the movement data and each UAV's residual energy level for guaranteeing high-level communication stability while forecasting a sudden link breakage before occurring. Using a strong route detection process, routing paths are explored to take into account the link breakage prediction, the balanced energy consumption, and the connectivity level of the explored pathways [13].

In [14], a data dissemination method is provided by constructing a virtual topology based on the charge of WSN nodes using software-defined networks (SDNs) via UAVs. Constantly, the topology is monitored and reconfigured if necessary. For facilitating simultaneous communication with the ground nodes, the SDN controller and the base station, the aerial nodes are armed with multiple-input multiple-output (MIMO) antennas. Within the proposed method, an efficient sleep timer and back-off counter approaches are used as well. The topology formation and preservation of a sleep timer and a back-off counter are facilitated by the SDN controller.

The problem is intensified by a sporadic network connection disrupting communication in UAVs. Therefore, a drone requires a deep learning-based, adaptive Intrusion Detection System to recognize its intruders and guarantee its safe return-to-home (RTH). In the suggested IDS, using Self-Taught Learning (STL) with a multiclass SVM, the IDS's high true positive rate is maintained, even in unknown territory. The Deep-Q Network is used by the self-healing technique in the IDS recovery phase that is a deep reinforcement learning algorithm for dynamic route learning facilitating the safe return home of the drone. Based on the simulation outcomes, the effectiveness of the proposed IDS is represented [15].

In [16], the UAV (physical layer security of an unmanned aerial vehicle) network is studied, in which the information is transmitted by a UAV-B (UAV base station) confidential to multiple information receivers (IRs) by assisting a UAV jammer (UAVJ) by existing the multiple eavesdroppers. Here, an optimization problem is formulated to mutually design the trajectories and convey the power of UAV-J and UAV-B for maximizing the minimum average secrecy rate overall IRs. The optimization problem is non-convex with the coupled optimization variables leading to the mathematically inflexible optimization problem. Hence, the optimization problem is decomposed into two subproblems and then solved using the succeeding convex approximation technique and an alternating iterative algorithm.

In [17], two aspects of secure communication and cooperative control are considered. The cooperative control is implemented by a clustering algorithm to increase the speed of converging the multi-UAV formation. Adjusting the flight control factor for accelerating the convergence of multi-UAV, a flock is created by the UAV group. For facilitating secure communication, the hierarchical virtual communication ring (HVCR) strategy is arranged to decrease the boundary of group communication and minimalize the insecure range.

In this paper, a method is proposed to maintain the security in UAV networks within surveillance, by verifying the data regarding events occurring from various sources. Hence, UAV networks are able to adapt peer-to-peer information stimulated by the blockchain principles and to discover the compromised UAVs in terms of trust policies. In the suggested method, secure asymmetric encryption is used with the official UAVs' pre-shared list. This method makes possible to detect the wrong information when hijacking an official UAV physically [18].

In [19], the authors suggest a method called SCOTRES, which uses five criteria to enhance network decision making and increase lifetime and load balance in the network. This method also considers the energy consumption of the nodes to provide cooperation between the nodes.

This paper investigates the trajectory design and resource allocating for energy-efficient secure unmanned aerial vehicle (UAV) communication systems in which multiple legitimate ground users are served by a UAV base station while existing a potential eavesdropper. Our objective is to maximize the UAV's energy efficiency while optimization of its user scheduling, transmit power, velocity, and trajectory. The formulation of the design is a nonconvex optimization problem considering the minimum data rate requirement of each user, the maximum tolerable signal-to-noise ratio (SNR) leakage, and the location ambiguity of the eavesdropper. To attain an efficient suboptimal solution, an iterative algorithm is suggested [20].

This paper studies a joint optimization problem of ground terminals (GTs) association under wiretap channels, unmanned aerial vehicle (UAV) flight trajectory, and downlink transmission power. Precisely, a scenario is considered, in which a group of GTs is served by a UAV and the minimum secrecy rate is maximized to guarantee the fairness among GTs. We establish an iterative algorithm in terms of the alternating and successive convex approximation (SCA) approaches for solving the nonconvex optimization problem [21].

Through unmanned aerial vehicles (UAVs), it is possible to support surveillance even in areas with no network infrastructure. By UAV networks, the security challenges are raised as a result of its dynamic topology. In the present study, a method is proposed to maintain the security in UAV networks within the framework of surveillance, by verifying data regarding events from various sources. Thus, UAV networks are able to adapt peer-to-peer general information stimulated by the blockchain ethics in terms of the trust policies. In this technique, secure asymmetric encryption is used with a pre-shared list of official UAVs. This work states detecting the misinformation when hijacking an official UAV physically [22].

In [23], an innovative trust model is proposed for UAVNs in terms of the mobility and performance pattern of UAV nodes and the features of inter-UAV channels. The suggested trust model includes 4 parts of the indirect trust section, the direct trust section, the trust update section, and the integrated trust section. According to the trust model, the perception of a secure link in UAVNs is formulated existing only a trust link and a physical link between two UAVs. Furthermore, the connectivity of UAVNs is analyzed by adapting the metrics of the secure connectivity probability and physical connectivity probability between two UAVs. Utilizing stochastic geometry with Doppler shift or without it, we originate analytical and accurate expressions of the secure connectivity probability and the physical connectivity probability.

In [24], a security model is suggested in terms of Identity Based (IB) authentication outline for UAV-integrated HetNets. the AVISPA tool is used to screen the absolutism of such a proposed scheme and some of its results indicated that our outline is resistant to the susceptibilities of intruders like replay, and impersonation.

Table 1 provides a complete overview of previous work in terms of; features and validation tools. All of the articles presented in this study have been published in famous journals. Also, these articles are acceptable in terms of citation level.

Table 1: Comparison between detection schemes for UAV

| Ref | Attack type | Features | Validation |
|---|---|---|---|
| [10] | Cyber attack | • Reduce overheads<br>• Solve privacy<br>• Increase confidentiality | NS-3 |
| [11] | Opportunistic attacker | • Solve security issues arise during M2M communication<br>• Achieves high level security | Mathematical |
| [12] | Cyber security threats | • Increase confidentiality<br>• Increase the detection rate | NS-3 |
| [13] | Flooding | • Upgrading the network lifetime parameter by minimizing energy consumption | NS-2 |
| [14] | Hybrid | • Securing long-term user's identity privacy<br>• Achieves high level security<br>• Increase of QoS | NS-3 |
| [15] | Jamming attacks | • Self-healing<br>• Improvement of accuracy, sensitivity and specificity | ONE simulator |
| [16] | Physical layer security | • Overcoming old key management that did not perform well.<br>• Identification and mutual authentication<br>• Achieves high level security | NUMERICAL |
| [17] | Hidden terminals | • Minimizes energy consumption<br>• Overcome the symmetric key distribution<br>• Identification and mutual authentication | MATLAB |
| [24] | Packet modification attacks | • Increase confidentiality<br>• Overcome the challenges of traditional public key | Avispa Tool |

## 3 Human Immune System (HIS)

HIS as the human's basic protection system supports human beings to survive diseases and environmental threats. Furthermore, by resembling the internet to humans in different ways, it is possible to develop an immune system for the internet in terms of the HIS's fundamentals. Immunity system denotes all bodily mechanisms in charge of protection of the body against detrimental agents in the situation like microorganisms and their products, pollen grains, drugs, and chemicals. The HIS includes three defensive lines operating in cooperation. Mucous, skin, secretions of skin, and membranes are included in the first layer. Phagocytic white blood cells, the inflammatory responses, and antimicrobial proteins are the subsections for the second layer. Ultimately, the third layer as the specific defensive mechanism involves antibodies and lymphocytes. Antibodies react to aberrant body cells, particular microorganisms, toxins and other materials signed by foreign molecules specifically. Two innate immunity system and acquired immunity system are included in the human immunity system [24].

### 3.1 HIS Algorithm

According to the former part, the HIS is a relatively complex mechanism able to protect the body against a tremendous group of irrelevant pathogens. It constructed mechanism of HIS is remarkably effective considerably in self and non-self-antigens distinction. The non-self-antigen is any external factor able to trigger an immune response like an attack or the bacteria. However, self-antigens are on the reverse side of non-self-antigens. The self-antigens are the cells of the living. Clonal selection, affinity, and negative selection are the main theories about HIS algorithms [24]. The pseudo-code for the Negative Selection Algorithm is confirmed in Algorithm 1.

---

**Algorithm 1:** Pseudo code for Negative Selection Algorithm

1: **Procedure Negative Selection Algorithm**
2: **Input:** A $S \subset U_i$ ("self-set"); a set $Mo \subset U_i$ ("monitor set"); an integer $ni$
3: **Output:** For each element $mo \in Mo$, either "normal UAV" or "malicious UAV".
4:     // phase I: Training
5:     $de \leftarrow$ empty set
6:     **while** $|De| < ni$ **do**
7:         $de \leftarrow$ the random detector set
8:         **if** $de$ does not match any element of $Si$ **then**
9:             **insert** $de$ into $De$
10:         **End if**
11:     **End while**
12:     // phase II: Classification
13:     **For each** $mo \in mo$ **do**
14:         **if** $mo$ matches any detector $de \in De$ **then**
15:             output "$mo$ is non-self" (an attacker)
16:         **else**
17:             output "$mo$ is self"
18:         **End if**
19:     **End For**
20s: **End Procedure**

---

## 4 The proposed ASP-UAVN approach

In this section, the proposed method, which is based on a method of self-adaptation and self-defense against lethal attacks, is examined in full detail. Six sections are included in the ASP-UAVN: in Sect, 4.1. The motion direction of the UAV is discussed. Sect 4.2. deals with the information exchange pattern of UAVNs. In Sect. 4.3 ASP-UAVN network model is discussed. In Sect, 4.4. the evaluation agent (to evaluate the routes) is discussed. Sect 4.5. deals with the decision-making agent, and in Sect. 4.6 defensive agent in ASP-UAVN is discussed.

### 4.1 ASP-UAVN network model

We take into account a UAS network where UAVs are arranged in an infinite 3D Euclidean space based on a homogeneous Poisson Point Process (PPP). A maximum one-hop communication range is included in the UAVs. A UAV is able to convey the data to the considered destination UAV straightly, or through a relay by one or further UAVs. A multi-hop outline is decode-and-forward, where an arriving packet is decoded by the relaying UAV then transmitted to the next hop. Moreover, a safe solution is presented in the ASP-UAVN network model, to protect the UAVs that are operative on two perspectives: First, it contains low false negative and positive rates and high detection accuracy. Second, it quickly discovers and separates attacks. In the suggested technique, the security issues like SF, WH, and SH attacks able to target the UAV are prohibited. Other properties should be added to Table 2 for detecting the attacks with high accurateness.

## 4.2 Information exchange pattern of UAS

The information is exchanged through the typical process. Originally, a message is delivered by a source ground station ($G_{src}$) to a UAV ($U_1$). Then, this UAV passes through a distance ($D_1$) to satisfy and send the message to another UAV ($U_2$). The message is delivered then by this UAV to another UAV ($U_3$) continuing in the same mode until delivering the message by the final UAV ($U_N$) to the ground station in the destination ($G_{DST}$). To minimize the latency in delivering end-to-end packets, each UAV in this procedure is directed to satisfy the next UAV exactly at the selected time.

However, in real environments as a result of the different performances of UAVs owing to changes in environmental uncertainties and engines, they fly at various velocities, hence, it is impracticable to anticipate that all UAVs can follow the same pattern. For example, UAVs with greater speeds may pass longer distances in comparison to others. Furthermore, it is possible to establish a communication line within two UAVs only into the communication range. Hence, it is essential to develop an association between UAVs. By the two UAVs in the communication range, or by their similar connection area, they will be able to exchange the information packets. This process needs a huge deal of time. No data packet exchange is probably happened by a UAV traveling in a connectionless area or outside the connection area.

## 4.3 Motion Direction of the UAV

To offer motion for the UAVs, in this work, the smooth turn (ST) mobility model was employed. ST makes the UAVs contain smoother trajectories such as taking turns with a larger radius or flying in straight trajectories. Thus, ST has a wide usage in analyzing UASs. This model can capture the UAVs' acceleration correlation in both spatial and temporal domains accommodating the analysis and design. Based on [27], a uniform distribution exists for the ST model's stationary node leading to some closed-form connectivity.

Table 2: Lethal attacks features

| Cyber security threats | Features |
| --- | --- |
| Wormhole attack | Deceiving UAVs in the process of discovering the route |
| Selective forwarding attack | Deceiving UAVs in the process of discovering the route |
| Sybil attck | Deceiving UAVs in the process of discovering the route |

All the notations and abbreviations used in this article (proposed method section and other sections) are given in Table 3.

Table 3: Acronyms and notations.

| Acronyms | Abbreviated acronyms | Notation | Abbreviated notations |
|---|---|---|---|
| $NS-3$ | Network Simulator 3 | $RREQ$ | Route Request |
| $NAM$ | Network Animator | $RREP$ | Route Reply |
| $IDS$ | Intrusion Detection System | $UAV_S$ | Source $UAV$ |
| $HIS$ | Human Immune System | $UAV_D$ | Destination $UAV$ |
| $AIS$ | Artificial Immune System | $ST$ | Smooth Turn |
| $ASP-UAVNs$ | Agent-Based Self-Protective Unmanned Aerial Vehicle Networks | $Th$ | Threshold |
| $UAV$ | Unmanned Aerial Vehicles | $P_m(r)$ | Probability $_{malicious}$ (route) |
| $GPS$ | Global Positioning System | $SSI$ | Signal Strength Intensity |
| $WH$ | Wormhole | $F_r$ | Fitness $_{route}$ |
| $SF$ | Selective Forwarding | $P_{UAV_M}$ | Probability Malicious UAV |
| $SH$ | Sinkhole | $MaxRTT$ | Maximum RTT |
| $FP$ | False positive | $SSI_i$ | Signal Strength Intensity $_i$ |
| $FN$ | False negative | $MaxSSI$ | Maximum SSI |
| $TP$ | True positive | $D$ | Distance |
| $TN$ | True negative | $R$ | Route |
| $DR$ | Detection rate | $AS$ | Antigen Self |
| $SFA$ | Security Framework Aircraft | $Ab$ | Anti-body |
| $DoS$ | Denial of Service | $UAVNs$ | Unmanned Aerial Vehicles Networks |
| $Ag$ | Anti-gen | $PLR$ | Packet Loss Rate |

**Employing Agents to Detect Attacking UAVs:** In our proposed method, the safest route from the starting point to the destination is chosen according to a self-matching system. In this method, a multi-agent system using an artificial immune system is employed to detect the attacking UAV and choose the safest route. In the proposed ASP-UAVN, the route request packet (RREQ) is initially transmitted from the source UAV to the destination UAV to detect the existing routes. Then, once the route response packet (RREP) is received, a self-protective method using agents and the knowledge base is employed to choose the safest route and detect the attacking UAVs. In ASP-UAVN, three types of agents are considered, including:

- Evaluation agent (to evaluate the routes)
- Decision making agent
- Defensive agent

These agents have modules distributed in different segments of the UAS, and each have a distinct responsibility. All agents are connected to the knowledge base to register the data and utilize the registered information. Figure 4 demonstrates the relationship between the agents and between the agents and the knowledge base. In the proposed method, agents are considered to detect Selective Forwarding attack, Wormhole attack, and Sybil attack.

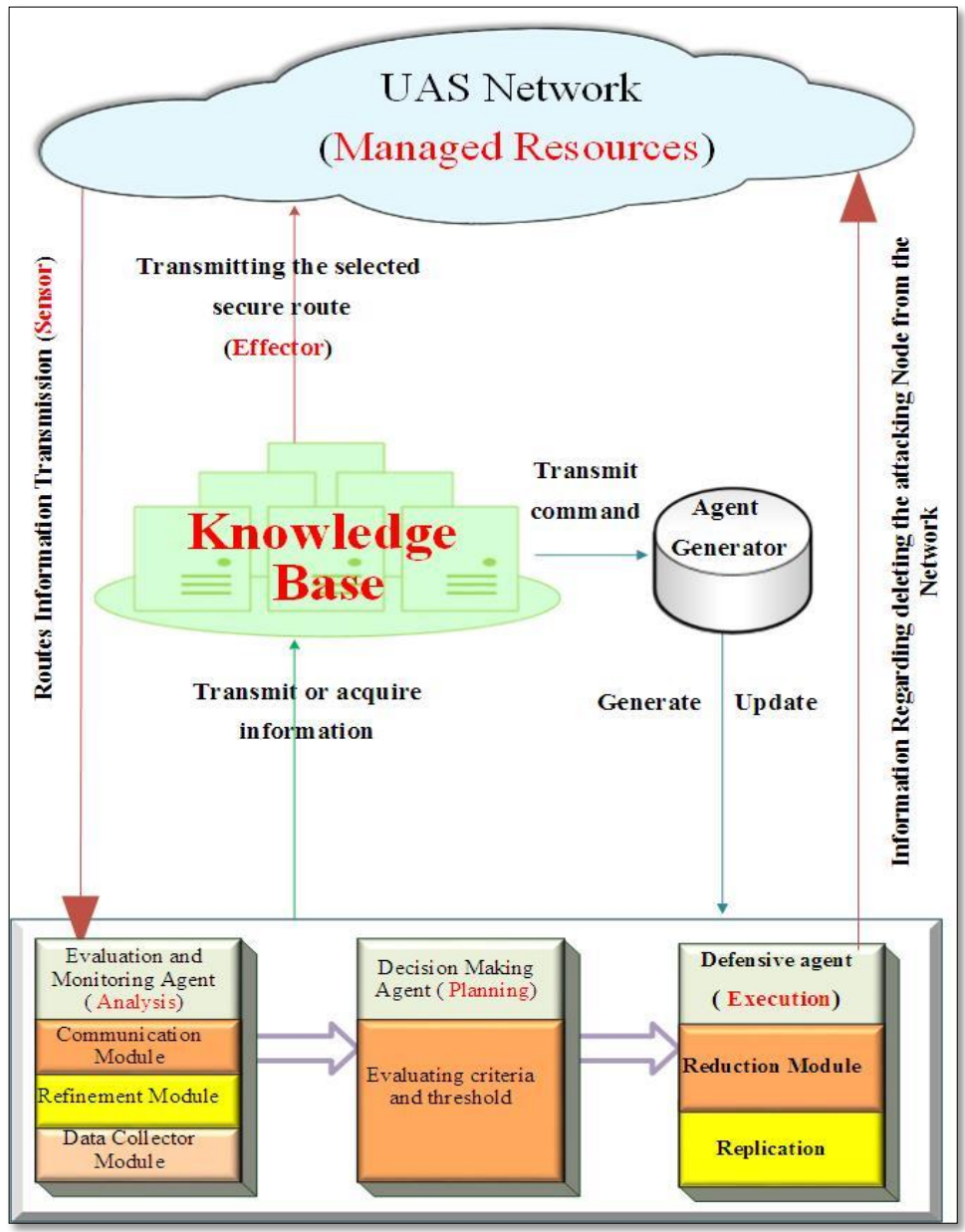

Figure 4: Relationship between agents with the knowledge base and UAV network

### 4.4 The evaluation agent

Antigens are considered as the set of all detected routes from the source UAV to the destination UAV. The considered evaluation agents are similar to the T-cells in the immune system: they are responsible for evaluating abnormal behavior of the existing UAVs in routes and reporting their behavior. These agents have three functional modules as follows:

1. Module for Data Collection
2. Module for Refinement
3. Module for Communication with the Decision-Making Agent

The evaluation agents in ASP-UAVN are agents that evaluate the existing routes to the destinations (i.e. the antigens); so that destructive behavior of the UAVs on each route can be detected.

The routes evaluation agent can be in the form of a table in the source UAV that registers the existing routes to the destination. For instance, the UAV in the source S has three routes to the destination, namely $R_1$, $R_2$, and $R_3$. The route evaluation agent is demonstrated in Table 4.

In this process, the routes for which the RREP message is sent to the destination UAV are tested in terms of a security. For this aim, one "Hello Packet" is transmitted over each route, and the destination UAV is responsible for transmitting a confirmation packet on the routes containing UAVs, following the receiving of the "Hello Packet". It is evident that if a route is contaminated with a destructive UAV, the "Hello Packet" will fail to reach the destination and thus, no confirmation packet will be received in the source. In such cases, the probability for the route to be contaminated increases (i.e. the value for $P_{UAV_M}(r)$ increases for the route r). On the other hand, if the "Hello Packet" reaches the destination, the confirmation packet will be received in the source, indicating that the route does not contain a destructive UAV (i.e. the value for $P_{UAV_M}(r)$ decreases). The procedure for transmitting the "Hello Packet" is repeated 4 times.

**The initial value for $P_{UAV_M}(r)$:** If the desired route is secure and has no malicious UAV, the $P_{UAV_M}(r)$ variable is set to zero. Conversely, if the route is not secure or in other words has a malicious UAV, this variable is set to 100. Next, to update the value for this variable, the source UAV transmits a "Hello Packet" to the destination UAV through all existing routes in its table in 4 iterations. If a confirmation packet is received from the destination UAV, 25 units is decreased from the value of $P_{UAV_M}(r)$. However, if no confirmation is received from the destination UAV, 15 units is added to the value for $P_{UAV_M}(r)$. This process is repeated 4 times during the route discovery process, and the value for $P_{UAV_M}(r)$ is updated for all routes. Finally, if the value for $P_{UAV_M}(r)$ is more than 50 for a route, it is rejected. The rest of the routes are sent to the decision-making agent and the knowledge base.

Table 4: Examining the attacker detection mechanism using the probability variable $P_{UAV_M}(r)$

| Response from the routes | Confirmation that packet is received | Confirmation that packet is not received | $P_{UAV_M}(r)$ |
|---|---|---|---|
| Response for Route 1 | √ | | $P_{UAV_M}(r) - 25$ |
| Response for Route 2 | | √ | $P_{UAV_M}(r) + 15$ |

The decision-making agent is capable of synthesizing the information on breaches to reach a precise decision regarding the breach.

### 4.5 The decision-making agent

The decision-making agent is similar to the B-Cells in the AIS, and is capable of making effective decisions regarding the distribution of the attacks. The main objective of the decision-making agent in the ASP-UAVN is detecting the existence of unfamiliar patterns in a potentially-large set of the existing familiar patterns.

Moreover, when this agent detects a suspicious route, it transmits the information regarding its set to the knowledge base instantly, so that the knowledge base can contact the agent generator to generate new agents to evaluate, make decision, and defend against these unknown attacks.

Making decision on the routes is carried out using 4 criteria, namely Delay, the Ratio for delivering healthy packets from the previous step ($PDR_H$), Packet Loss Ratio (PLR), and the Frequency of sending packages in each route (FSR), according to the considered attacks. For instance, in some attacks, the destructive UAV deletes all packets, while in some others, the invading UAV deletes only some of the packets.

Therefore, the decision-making agent examines these four criteria based on the information obtained during pre-acquisition and acquisition stages of every route. For suspicious routes, the decision-making agent determines the threshold and transmits a warning to the defensive agent so that it can detect the destructive UAVs. In Table 5, the considered equation for the threshold is in a way that the existence of high delay, high PLR, low ratio for receiving the Hello packet, and high number of repeats (i.e. transferring repeated packets) denotes the existence of a destructive UAV in the route.

Table 5: Decision making agent

| Suspicious routes | Delay | PLR | PDR | FSR |
|---|---|---|---|---|
| R₁ | 30ms | 15% | 85% | 24 |
| R₂ | 10ms | 25% | 75% | 3 |
| R₃ | 20ms | 5% | 95% | 2 |
| Threshold | $Th = \left( \dfrac{Delay}{MaxDelay} + \dfrac{PLR}{MaxPLR} + \dfrac{MaxPDR}{PDR} + \dfrac{FSR}{MaxFSR} \right)$ | | | (1) |

$P_{DR}(r)$: The value for the threshold (Th) for each route is determined according to the four criteria mentioned, and is registered in $P_{DR}(r)$. The route with the highest threshold is eliminated and is sent to the defensive agent.

### 4.6 Defensive agent

Defensive agents act similar to the antibodies exuded by the lymphocyte. Their functional modules include replication and reduction modules. Defensive agents can evaluate the existing UAVs in the defective route to perform proper actions according to the information provided by the decision-making agent. According to the procedure, when an attacker UAV is detected, the neighboring UAVs are requested not to resend the packets they received from the attacker UAVs.

To detect the attacker in the designated route, the agents replicate themselves in the vicinity of each node and send some Test packets over the route in consecutive periods. A test packet is a packet similar to the normal packets in the UAS network. Therefore, the attacking UAV receives and tries to delete it. The defensive agent detects the destructive UAV in a suspicious route using the following Eq. (2):

$$\begin{cases} M_1 = X_1 & For\ t = 1 \\ M_t = \alpha * X_t + (1-\alpha) * M_{t-1} & For\ t > 1 \end{cases} \qquad (2)$$

Where $\alpha$ is the regulation factor coefficient with a constant value between 0 and 1, $X_t$ is the value for the test packet in a time interval $t$, and $M_t$ is the mean transmission value for the test packets in every time interval $t$.

In every interval, a predetermined number of test packets are transmitted in the designated route. Then, for each UAV, the defensive agent determines the number of the test packets transmitted by that specific UAV. If the total number of the test packets transmitted by a UAV is fewer or equal to the value for $M_t$, it demonstrates that this UAV is a destructive UAV that removes a number of the packets. The value for the coefficient $\alpha$ is considered a constant value between 0 and 1. Considering lesser value for this coefficient indicates higher expected probability for the loss of packets. On the other hand, if the value for $\alpha$ is considered closer to 1, it indicates that we expect fewer packets to be lost.

Using this method, the defensive agents can successfully isolate the attacking UAVs, as illustrated in Figure 5. To remove the breaches, the defensive agents replicate themselves and barricade further replication of the attacker after a certain time interval by removing the attackers.

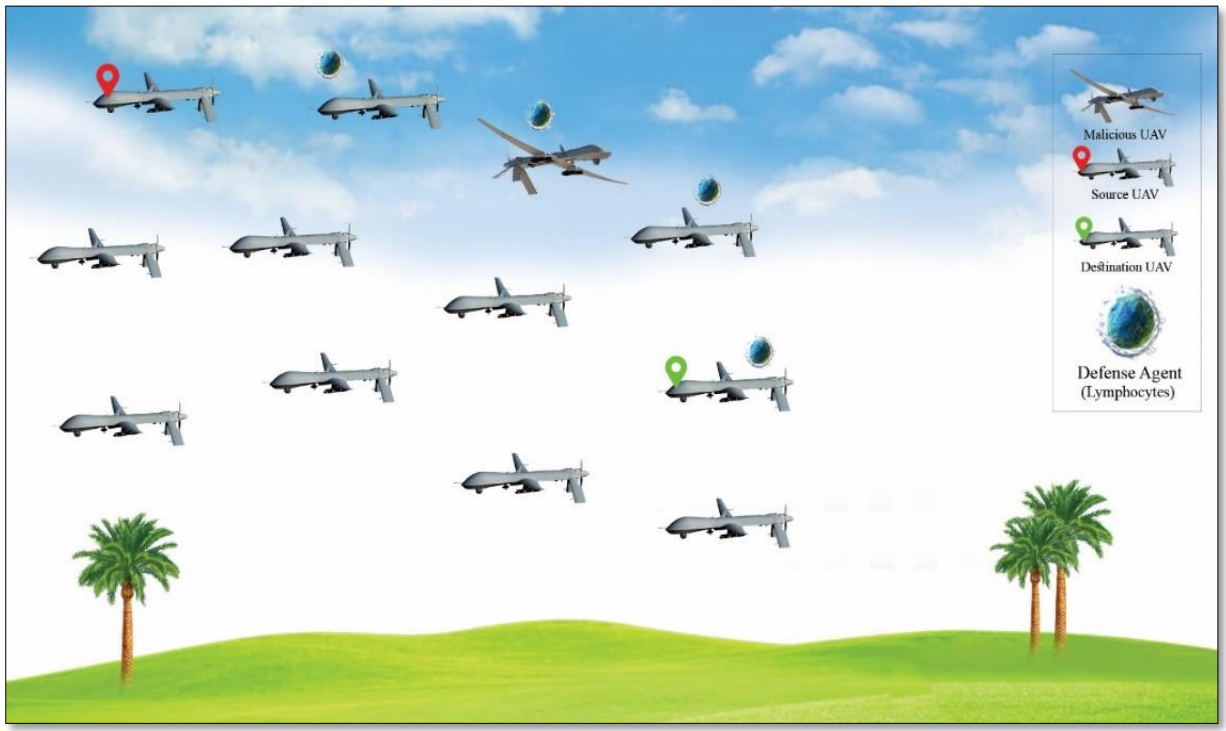

Figure 5: Defensive agents replicate themselves to evaluate and detect the destructive UAVs in suspicious routes.

When the defensive agent detects a destructive UAV, it will ask all the UAVs in the route to disregard and delete any packets received from this UAV. In addition, it sends a message regarding detection of the destructive UAV to the knowledge base, so that the destructive UAV is no longer employed for route finding.

**The Knowledge Base Layer:** The knowledge base is in connection with all agents, and they needed to be intelligently evolved to defend against a wide range of attacks. The knowledge base includes the following stages to choose the most secure route (based on the information received from the agents) for transmission of data in the immune memory:

**Affinity:** as defined in the immune system of the human body, the main objective of B-Cells is creating antibodies against antigens, and ultimately evolving into memory cells once they are activated by antigen interactions. The memory cells generate more antibodies in shorter time during further impacts with the same antigen. In our proposed method, to choose the best B-cell and to perform affinity, routes with low latency, low ratio for loss of packets, high ratio for receiving packets, and low packet re-transmission number are chosen. In other words, at this stage, routes with low threshold value are chosen.

**Matching:** in this stage, routes with low threshold are compared and evaluated using the following two attributes to choose the safest route. In addition, the most significant attribute of the attacker-detection mechanism is its capability to be corrected over time. In other words, they need to be correctable and have the capacity for easy learning.

The details for the first and second attribute of matching: the details are described in the following two attributes:

**Attribute 1, round trip time between the source UAV and the destination UAV:** the knowledge base calculates the round-trip time for all the received routes from the source UAV to the destination UAV based on the acquired information.

**Attribute 2, the signal strength index of the received signal:** the invader generates a high strength signal (the SSI that a destructive UAV has generated) to gain control over the UAV. The procedure for detection is that the knowledge base initially collects all the generated SSIs from the transmitters. It then compares these values with the value for a normal SSI (i.e. SSI generated by a normal UAV). In this manner, suspicious generated signal strengths and normal ones are distinguished.

**Finalizing the Detection Set:** for all the routes with low threshold (Th) value, the safest route is selected according to the Algorithm 1 demonstrated in Figure 6:

**Algorithm 1:** Pseudo code for ASP-UAVN approach

1: **Initialize** the Antigen collection time to 15s;
2: **Initialize** the Antigen towards min to 80s;
3: **Initialize** the Delay buffer size max to 1400;
4: **Initialize** the Storing time to 10s;
5: **Initialize** the Max number of antigens to 1000;
6: **Let** $P_{DR}(r)$ as the probability for destructiveness of the route
7: **Let** $N$ represents the number of candidate routes between the source UAV and the destination UAV
8: **Let** $F_r$ represents the fitness route
9: **Let** $UAV_{PR}$ represents a secure and reliable route between UAVs
10: **Procedure** Selecting a safe and reliable route
11:     For $r = 1$ To $N$ Do
12:         Calculate the value for $Th$ in every route
13:         IF $Max_{(Th)}(r) < P_m(r)$ **Then**
14:             Remove the route and send a warning to the defensive agent
15:         **Else**
16:             Select routes with the lowest $Th$ value $Min(Th)$
17:             Calculate the value for $F_r$ according to this equation:
18: $$F_r(RTT_i, SSI_i) = \left(\frac{MaxRTT}{RTT_i}\right) + \left(\frac{SSI_i}{MaxSSI}\right)$$
19:             The route with the following criteria is selected:
20:             $UAV_{PR} = Min(Th) \& Max(F_r)$
21:         EndIf
22:     EndFor
23: **End Procedure**

Figure 6: The algorithm for selecting a secure and reliable route

According to Algorithm 1, once the threshold value ($Th$) for each route was determined and the destructive routes were refined, the routes with low Th value are once again compared according to the function $F_r$ for the fitness route, and the route with the highest $F_r$ is selected as the safest route.

**Hyper-mutation:** among the evaluated routes, those with approximately similar conditions (i.e. minimum threshold level and maximum evaluation function) are transferred into the hyper-mutation stage. At this stage, the routes are evaluated with another criterion (i.e. in a similar condition, the route with the highest PDR is selected), so that the safest route for the UAVs is selected.

**Registration in the Security Memory:** the routes that meet the conditions of the equation $UAV_{RR}$, or those that have the highest PDR value following the hyper-mutation stage, are the safest routes. Therefore, they will be registered in the security memory for further use.

The flowchart for the proposed ASP-UAVN is demonstrated in Figure 7.

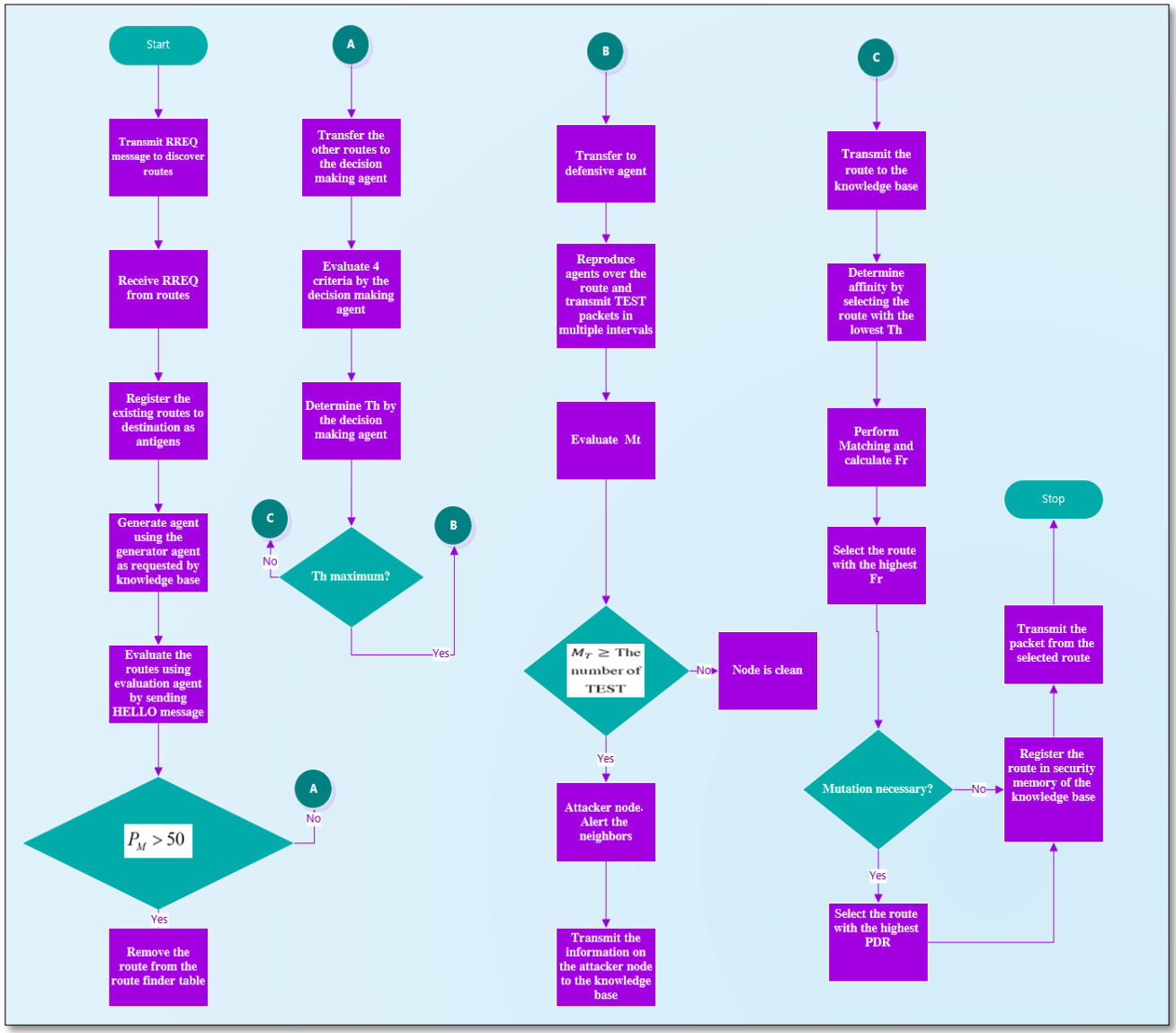

Figure 7: Flowchart of the ASP-UAVN

**Analyzing the characteristics of ASP-UAVN**: There are numerous advantageous characteristics in the proposed ASP-UAVN by employing artificial immune system and multi-agent intelligent technology. In the following, a number of these characteristics are discussed.

**Distributivity:** As our considered agents are distributed in all UAVs similar to the distribution of lymphocytes in the body, and since the knowledge base is updated periodically or when the UAV is under attack, the three considered agents are logically independent, with intermediaries to enable communication. The evaluation agents are capable of controlling the performance of the UAVs on the route via transmitting HELLO messages. Moreover, the decision-making agents analyze the performance of the UAVs over the route based on the considered criteria to detect destructive routes. Furthermore, defensive agents can detect attackers independently and remove them from the network.

**Independence:** Similar to the body immune system that does not require external administration and maintenance for classification and elimination of the pathogenic agents, the knowledge base and the agents can evaluate, make decision, and defend against the attacker UAVs in the three considered attacks cooperatively (with the cooperation of other UAVs) or independently. In ASP-UAVN, the knowledge base and the agents can be updated or reproduced independently.

**Self-protection:** Similar to the body immune system that is capable of learning to defend against new pathogens and detecting the known pathogenic agents through the use of immune memory, the proposed ASP-UAVN is capable of detecting these attacks via a cooperation between the considered agents and the use of the knowledge base.

The proposed ASP-UAVN is a self-matching method that all the stages of self-matching, as illustrated in Figure 5, are applied to it according to the following steps:

**Administered Source:** the source in this project is the UAS network.

**Sensor:** Collecting UAV information via transmitting RREQ and receiving RREP in the UAS network.

**Evaluation:** This component collects the UAV information by employing the received RREPs. This activity is carried out using the data collection module in the evaluation agent.

**Analysis:** This component receives the information between the UAVs from the evaluation component. It then analyzes the acquired information to determine the desired routes. Using refinement and communication modules from the evaluation agent, this stage carries out its responsibility by refining the detected routes and discarding a number of them, followed by transmitting the desired routes to the decision-making agent.

**Planning:** this component makes decision regarding the safest routes. The planning for this component in the proposed ASP-UAVN is performed according to the decision-making agent by evaluating the four considered criteria and determining the value for the threshold (Th).

**Execution:** This component provides mechanisms for the network to perform planning. In the proposed ASP-UAVN, execution is performed in two stages, namely defensive agent (to delete the destructive UAVs), and knowledge base (to select the safest route via affinity and matching).

**Knowledge:** The common knowledge in the self-matching structure is the knowledge base employed according to the Figure 1 in the proposed method.

**Effector:** This component is employed to apply the final decision to the environment. In the proposed method, it is carried out via transmitting the selected safe route to the UAS by the knowledge base.

The proposed ASP-UAVN is effective and efficient in defense against Selective Forwarding, Wormhole, and Sybil attacks. ASP-UAVN is capable of effectively detecting these attacks in cooperation with the agents. In addition, in normal conditions, only a limited number of agents exist in the UAV. However, agents can reproduce and increase swiftly when needed, and decrease once the attacker is detected. In addition, the following two methods are considered in the proposed method to keep the knowledge base up to date:

- Periodic polling from all agents regarding the collection of abnormal behaviors of the network and operational performance of the agents.

- Active registration of information in the knowledge base by the agents if necessary (e.g. when an unknown attack occurs in large scale).

## 5 Theoretical analysis of ASP-UAVN

In this section, the proposed method is evaluated using theoretical analysis in two phases. The first phase involves the theoretical security analysis of the ASP-UAVN. In the second phase, the ASP-UAVN method is analyzed in terms of complexity and packet transmission.

### 5.1 Theoretical analysis of security

The theoretical security analysis of the proposed ASP-UAVN method and its effectiveness against malicious UAVs will be presented in detail in this section. The main part of theoretical security analysis is shown in Theorem 1. This describes the number of packets deleted by malicious UAVs. Therefore, according to this theorem, if malicious UAVs are identified and discarded from network data transmission, a secure connection will be obtained among the UAVs in the network. Therefore, we first provide definitions and then theorems and proofs.

***Definition 1.*** The variable $packet_{received}$ shows packets that have been successfully transmitted to the destination UAV.

***Definition 2.*** The variable $packet_{dropped}$ shows packets that were not successfully transmitted to the destination, i.e., were removed by malicious UAVs.

***Definition 3.*** The variable $TotalPacket$ shows the sum of $packet_{received}$ and $packet_{dropped}$.

***Definition 4.*** The variable $PDR$ shows the rate of $packet_{received}$ divided by $TotalPacket$.

***Theorem 1.*** The conditional expression $(packet_{dropped} - PDR)*(packet_{received}) \leq 0$ indicates an ideal UAV. The number of missing packages ($\eta$) can be neglected. The number of missing packets is a fraction of the number of packets transmitted. In particular, an upper limit for $\eta$ can be defined as below:

$$((packet_{dropped} - PDR)* packet_{received}) \leq \eta \qquad (3)$$

***Proof:*** In general, suppose there are $Total\_N$ UAV nodes in the UAV network where $Total\_M$ of them are malicious so that the condition $Total\_M < Total\_N$ is met. $(Total\_M * L)$ is sum of the links controlled by malicious UAVs. The maximum value for the $(Total\_M * L)$ is equal to $(Total\_M * N)$.

For example, consider a malicious link ($Fault_{link}$) in a UAV network which has been repeatedly reported as a malicious link for $c_{Fault_{link}}$ times while changing to a normal UAV for $r_{Fault_{link}}$ times. The maximum weight of the link $w_{Fault_{link}}$ is equal to $(Total\_M * len)$ where $(Total\_M * len)$ is the maximum length of an error-free route in the UAV network. If the weight of the link reaches to

$(Total\_M * len)$, its efficiency is considered lower than that of the non-defected routes. Therefore, $w_{Fault_{link}}$ can be calculated according to the equation below:

$$w_{Fault_{link}} = 2^{\left(c_{Fault_{link}} - r_{Fault_{link}}\right)} \tag{4}$$

The minimum number of times it is reported as a broken link is $\left(\dfrac{packet_{dropped}}{\gamma}\right)$ where $\gamma$ is the number of packets deleted by malicious UAVs, which is indicated by a defected route between the UAVs. Therefore, according to Eq. (5):

$$\left(\left(\dfrac{packet_{dropped}}{\lambda}\right) - \sum_{Fault_{link} \in (Total\_M*L)} c_{Fault_{link}}\right) < 0 \tag{5}$$

Where $\lambda$ is average deviation. Similarly, the maximum number of times it has reverted to a normal UAV is equal to $\left(\dfrac{packet_{received}}{\dfrac{\gamma}{TotalPacket}}\right)$. Therefore, according to Eq. (6):

$$\left(\sum_{Fault_{link} \in (Total\_M*L)} r_{Fault_{link}}\right) - \left(\dfrac{packet_{received}}{\left(\dfrac{\gamma}{PDR}\right)}\right) < 0 \tag{6}$$

Considering Eq. (7), we have:

$$\dfrac{\left(\dfrac{packet_{dropped}}{\gamma}\right) - \left(\dfrac{packet_{received}}{\left(\dfrac{\gamma}{PDR}\right)}\right)}{\sum\limits_{Fault_{link} \in (Total\_M*L)} \left(c_{Fault_{link}} - r_{Fault_{link}}\right)} \leq \tag{7}$$

According to Equation (5), $\left(c_{Fault_{link}} - f_{Fault_{link}}\right) \leq \log w_{Fault_{link}}$ and therefore:

$$\sum_{Fault_{link} \in (Total\_M*L)} \left(c_{Fault_{link}} - r_{Fault_{link}}\right) = \sum_{Fault_{link} \in (Total\_M*L)} \left(\log w_{Fault_{link}}\right) \tag{8}$$

By combining Eq. (7) and Eq. (8), we have:

$$\left(\left(\left(packet_{dropped}\right) - PDR\right) * packet_{received}\right) \leq$$
$$\left(\gamma * \sum_{Fault_{link} \in (Total\_M * L)} \left(\log w_{Fault_{link}}\right)\right) \leq \left(\gamma * Total\_M * Total\_N * \log(Total\_M * len)\right) \quad (9)$$

Where $\gamma = (b * \log(Total\_M * len))$ and $\gamma$ is the number of packets deleted by the malicious UAV in each operation.

$$\left(\left(packet_{dropped} - PDR\right) * packet_{received}\right) \leq \quad (10)$$
$$\left(\beta * Total\_M * Total\_N * \log^2(Total\_M * len)\right)$$

Therefore, the amount of malicious UAV intrusions can be limited in the network. The ideal situation is when there is no malicious UAV in the network, where $\left(packet_{dropped} - TotalPacket\right) * \left(packet_{received}\right) \leq 0$. If the number of malicious UAVs exceeds a threshold, they will be quickly identified and discarded from packet transmission between the UAVs using Eq. (11):

$$(packet_{dropped} \geq mal_{thr}) \rightarrow \quad (11)$$
$$(Total\_M = Total\_M - 1)$$

The rate of intrusion decreases by decreasing $(Total\_M * Total\_N)$. If all the malicious UAVs are detected, $(Total\_M * Total\_N)$ will be zero which is ideal. Therefore, when evaluating the forwarding performance of the packages as well as routing operations in the UAV network, this mechanism will protect the network against wormhole, selective forwarding and sink hole attacks. As a result, the communication among all the UAVs in the network will be secure.

### 5.2 Theoretical analysis of the complexity of time and message

In this section, we estimate costs in terms of the number of packets exchanged and the complexity of time and message. The meaning of acronyms and notations used in the equations are given in Table 6.

Table 6: Major acronyms and notations used.

| Notations | Means |
| --- | --- |
| *Range* | Communication range between UAVs |
| *b* | Messages between $UAV_{source}$ and $UAV_{destination}$ |
| *x* | Distance |
| *n* | Number of UAVs |
| *N* | Neighbor |

**Time complexity:** The time complexity is equal to the total time complexity of the RREQ messages and forwarded RREPs, and the packets delivered. Eq. (12) demonstrates $TimeComplexity_{ASP-UAVN}$.

$$TimeComplexity_{ASP-UAVN} = \left(TimeComplexity_{RREQ} * 3\right) \quad (12)$$

Where,

$$TimeComplexity_{RREQ} = \left(N_{Hops} * Delay_{UAVi,UAVI+1}\right)$$

$$N_{Hops} = \left\lfloor \frac{X}{Range} \right\rfloor$$

$Delay_{UAVi,UAVI+1}$: It takes time for the route to move between the UAVs that made the connection.

**Messages complexity:** The message complexity is equal to the total time complexity of the RREQ messages and forwarded RREPs, and the packets delivered. Eq. (13) demonstrates $MessageComplexity_{ASP-UAVN}$.

$$MessageComplexity_{ASP-UAVN} = \left(2*b + (n-1)\right) \quad (13)$$

Where,

$$b = \left\lfloor \frac{x}{Range} \right\rfloor + 1$$

Finally, the complexity of the route discovery process is equal to the re-forwarding of the RREQ package by the UAV other than the $UAV_{destination}$ that is a complexity of $\theta(\log n)$. If we consider the entire UAV network as a single, independent system, the complexity will be $\theta(n)$.

## 6 Evaluating the Performance

The ASP-UAVN performance is assessed in the following section to avoid the lethal attacks.

### 6.1 Performance metrics

Here, the performance and effectiveness of our suggested ASP-UAVN method are systematically assessed with complete simulations. A comparison is performed between the results and with SFA, BRUIDS, and SUAS methods proposed in [9], [10] and [11], respectively. The PDR, PLR, false negative, false positive, and detection ratio are assessed. The meaning of notations used in the equations are given in Table 7.

Table 7: The parameters specified for PDR and PLR

| Notations | Means |
| --- | --- |
| $X_i$ | Number of packets received by node i |
| $Y_i$ | Number of packets sent by node i |
| n | Experiments |

**PDR:** This criterion represents the rate of packets that were successfully delivered to the destination [25, 26].

$$PDR = \left(\frac{1}{n}\right) * \left(\frac{\sum_{i=1}^{n} X_i}{\sum_{i=1}^{n} Y_i}\right) * 100 \tag{14}$$

**PLR:** This measure represents the percentage of packets deleted by malicious UAVs. The higher the percentage, the worse the performance of the method, but on the contrary, if the percentage of this variable is low, it shows the efficiency and excellent performance of the method [27, 28]. The rate of packets removed is calculated by Eq. (15).

$$PLR = \left(\frac{1}{n}\right) * \left(\frac{\sum_{i=1}^{n} Y_i - \sum_{i=1}^{n} X_i}{\sum_{i=1}^{n} Y_i}\right) * 100 \tag{15}$$

**FP:** The FP is determined by the total number of UAVs mistakenly found as the intruder UAVs divided by the total number of normal UAVs [29, 30]. Hence, Eq. (16) illustrates the

$$FP = \left(\frac{FP}{FP + TN}\right) * 100 \quad \text{Where:} \quad TN = \left(\frac{TN}{TN + FP}\right) * 100 \tag{16}$$

**FN:** The rate of the intruder UAV to total normal UAVs incorrectly signed as a normal UAV [31]. The calculation is proved by Eq. (17).

$$FN = \left(\frac{TP + TN}{All}\right) * 100 \quad \text{Where:} \quad TP = \left(\frac{TP}{TP + FN}\right) * 100 \tag{17}$$

**DR:** Ratio of intruder nodes to total lethal attacks that were correctly diagnosed as intruder attack. Eq. (18) determines the DR [32].

$$DR = \left(\frac{TP}{TP + FN}\right) * 100 \quad \text{where} \quad All = TP + TN + FP + FN \tag{18}$$

**Antibody definition:** An antibody is a kind of protein that is generated in the immune system to answer the existence of a special antigen. It circulates in the blood or remains in the generating location to attack the antigen (usually, the foreign object like a bacteria, a virus, even a normal body tissue, or a food material), and makes it harmless. Each antibody recognizes a special antigen as its target. The antibodies are secreted from the Plasmocytes, and they are related to Humoral immunity. After the collision of lymphocyte B with its special antigen, it is separated, and it creates a B memory cell and the plasmocyte. The memory cells are always in ambush to divide rapidly when a re-collision with that antigen. The plasmocytes don't have any antigen receivers, and their duty is antibody generation. Moreover, they have a rough endoplasmic reticulum and an extensive Golgi body. The antibodies are generated actively and entered into the blood. An antibody is soluble in blood, so it called Humoral immunity. (Humors: body fluids including lymph blood and interstitial fluid) antibodies have an antigen receiver similar to B lymphocyte and B memory that is the antigen complementary. The antibody doesn't destroy the antigen, but it neutralized it. It sticks to the antigen most simply and prevents its connection to the host cell. In this state, in addition to antigen inactivation, it makes easy macrophage phagocytosis. The antibodies affect the creation of asthma, allergies, and autoimmunity. It means that inappropriate generation of them results in these diseases. Also, it should not be misunderstood that the antibodies are not useful. Because without the antibodies, many diseases kick the human out. An antibody has a low effect on fighting cancer and transplant cells.

**6.2 Simulation setup and comparing algorithms**

Since implementing and debugging UAVNs in real networks is difficult, considering simulations as a basic design instrument is necessary. The primary benefit of simulation is simplification of analysis and verification of protocol, especially in large systems [33-35]. In this part, the suggested method's performance is assessed by NS-3 as the simulation instrument, and then the results will be discussed. It should be noted that all ASP-UAVN, SFA, BRUIDS, and SUAS settings and parameters are considered as equal.

**6.3 Simulation results and Analysis**

The ASP-UAVN performance is analyzed in this section under the four scenarios (Table 8). Initially, 500 nodes are deployed in the UAV area in a uniform manner. Table 8 gives some major parameters.

Table 8: Setting of simulation parameters.

| Parameters | Value |
| --- | --- |
| Channel type | Channel/Wireless channel |
| MAC Layer | MAC/802.11. b |
| Traffic type | CBR |
| UAV speed | 180 m/s |
| Transmission layer | UDP |
| Packet size | 512 Byte |
| Malicious UAV | 5%, 10%, 15% |
| Type of attacks | WH, SF, SH |
| Transmission range | 30 M |
| Selection of target UAV | Random |

Table 9 shows the important parameters used in all four scenarios. In this table, the percentage of destructive UAVs is 5, 10, 15 and 20%.

Table 9: The setting of simulation parameters for four scenarios.

| Scenario #1 | | Scenario #2 | |
|---|---|---|---|
| Number of Antibody | 200 | Number of Antibody | 200 |
| Malicious UAV rate | 5% | Malicious UAV rate | 10% |
| Topology | 1000 x 1000 | Topology | 2000 x 2000 |
| Time | 1000 | Time | 1000 |
| Scenario #3 | | Scenario #4 | |
| Number of Antibody | 200 | Number of Antibody | 50, 100, 150, 200, 250, 300, 350 |
| Malicious UAV rate | 15% | Malicious UAV rate | 20% |
| Topology | 3000 x 3000 | Topology | 4000 x 4000 |
| Time | 1000 | Time | 1000 |

In Table 10-14, the proposed method is compared with all three methods in all criteria. The number of antibodies is defined as 50 to 350. By increasing the number of antibodies, the results of each method are given in the following tables. As can be seen from the tables, the proposed method has performed very well.

Table 10: *PDR* (in %) vs number of antibodies.

| | PDR (%) | | | |
|---|---|---|---|---|
| Number of Antibody | BRUIDS | SFA | SUAS-HIS | ASP-UAVN |
| 50 | 37 | 38 | 57 | 82 |
| 100 | 38 | 40 | 61 | 83 |
| 150 | 40 | 42 | 65 | 84 |
| 200 | 44 | 46 | 59 | 87 |
| 250 | 42 | 48 | 56 | 88 |
| 300 | 40 | 51 | 64 | 90 |
| 350 | 46 | 55 | 70 | 91 |

Table 11: *PLR* (in %) vs number of antibodies.

| | PLR (%) | | | |
|---|---|---|---|---|
| Number of Antibody | BRUIDS | SFA | SUAS-HIS | ASP-UAVN |
| 50 | 60 | 58 | 41 | 16 |
| 100 | 57 | 53 | 38 | 14 |
| 150 | 55 | 48 | 37 | 13 |
| 200 | 50 | 43 | 35 | 11 |
| 250 | 48 | 40 | 34 | 10 |
| 300 | 45 | 39 | 32 | 9 |
| 350 | 41 | 36 | 30 | 8.5 |

Table 12: FP (in %) vs number of antibodies.

| Number of Antibody | FPR (%) | | | |
| --- | --- | --- | --- | --- |
| | BRUIDS | SFA | SUAS-HIS | ASP-UAVN |
| 50 | 0.083 | 0.069 | 0.059 | 0.037 |
| 100 | 0.078 | 0.061 | 0.057 | 0.032 |
| 150 | 0.064 | 0.052 | 0.048 | 0.03 |
| 200 | 0.059 | 0.052 | 0.042 | 0.028 |
| 250 | 0.045 | 0.032 | 0.029 | 0.024 |
| 300 | 0.038 | 0.03 | 0.025 | 0.02 |
| 350 | 0.021 | 0.026 | 0.018 | 0.019 |

Table 13: FN (in %) vs number of antibodies.

| Number of Antibody | FNR (%) | | | |
| --- | --- | --- | --- | --- |
| | BRUIDS | SFA | SUAS-HIS | ASP-UAVN |
| 50 | 0.119 | 0.08 | 0.07 | 0.065 |
| 100 | 0.102 | 0.077 | 0.067 | 0.0495 |
| 150 | 0.1 | 0.071 | 0.061 | 0.055 |
| 200 | 0.09 | 0.065 | 0.055 | 0.0485 |
| 250 | 0.08 | 0.06 | 0.06 | 0.0465 |
| 300 | 0.06 | 0.044 | 0.054 | 0.0425 |
| 350 | 0.06 | 0.049 | 0.052 | 0.0385 |

Table 14: DR (in %) vs number of antibodies.

| Number of Antibody | DR (%) | | | |
| --- | --- | --- | --- | --- |
| | BRUIDS | SFA | SUAS-HIS | ASP-UAVN |
| 50 | 71 | 68 | 75 | 81 |
| 100 | 72 | 71 | 76 | 82 |
| 150 | 73 | 74 | 77 | 84 |
| 200 | 74 | 76 | 78 | 88 |
| 250 | 75 | 77 | 79 | 90 |
| 300 | 76 | 78 | 80 | 91 |
| 350 | 77 | 79 | 81 | 92 |

**PDR:** Figure 8 compares the proposed method in terms of packet delivery rate with all three methods mentioned. The proposed method has a high PDR due to the ability to quickly detect lethal attacks on a self-adaptive basis. Because it detects attacks quickly and does not allow these attacks to remove exchanges between UAVs. Because this is prevented, it makes sense to increase the package delivery rate on the network. As shown in the figure below, in the worst-case scenario, when the percentage of lethal attacks is equal to 20%, the efficiency of the proposed method, SUAS-HIS, SFA and BRUIDS are equal to 90, 40, 50, and 45%, respectively. Hence, the excellent performance of the ASP-UAVN method is achieved due to the efficient cooperation between the decision-maker and the defense agent. These modules mimic the misbehavior of a malicious UAV and use a law-based, and self-defense detection to detect the intrusion of a malicious UAVs.

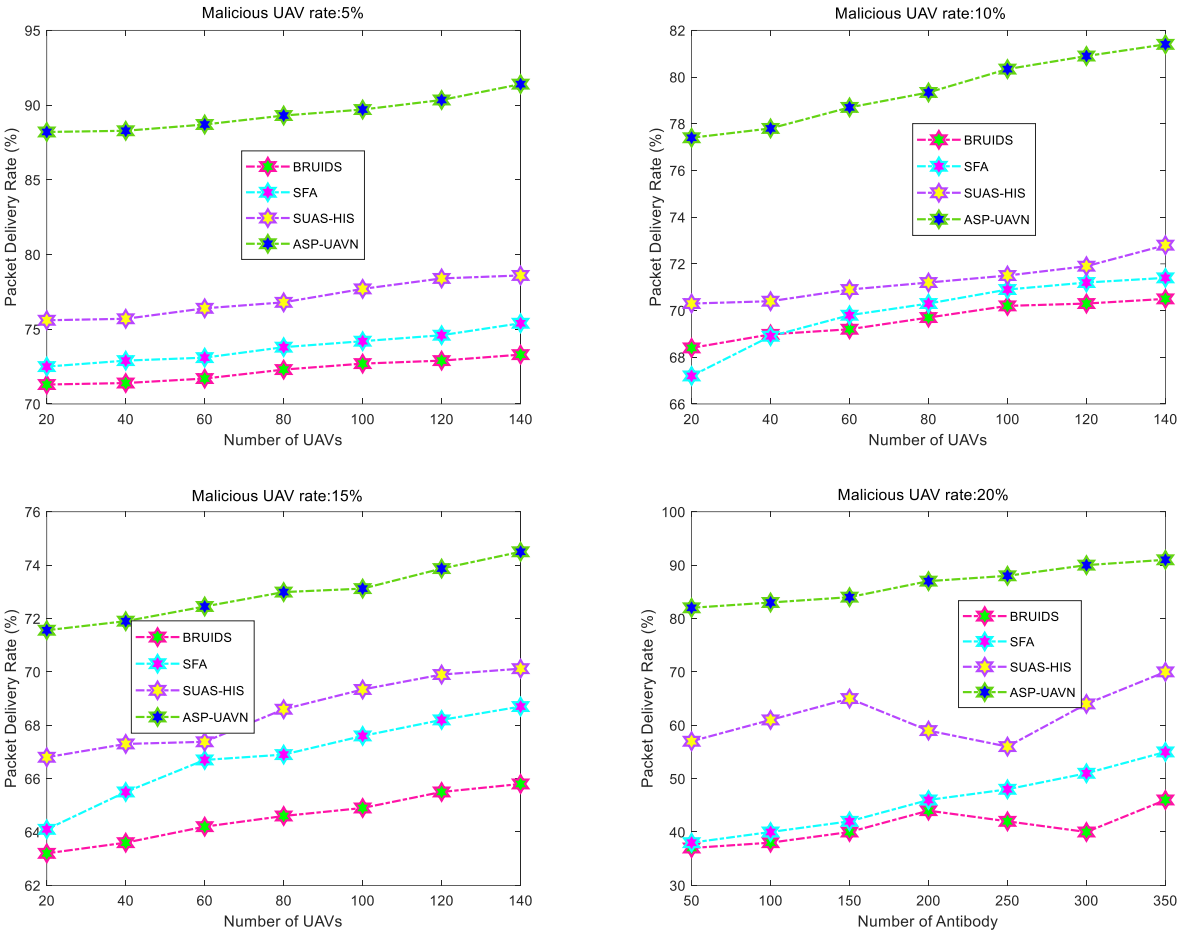

**Fig. 8** Comparison of the ASP-UAVN, SUAS-HIS, SFA and BRUIDS models in term of PDR.

**PLR:** Figure 9 compares the performance of ASP-UAVN with that of SUAS-HIS, SFA, and BRUIDS for detection of the lethal attacks. As shown in the figure, ASP-UAVN decreases the packet loss rate by more than 10.45, 17.54% and 27.05% those of SUAS-HIS, SFA, and BRUIDS, respectively.

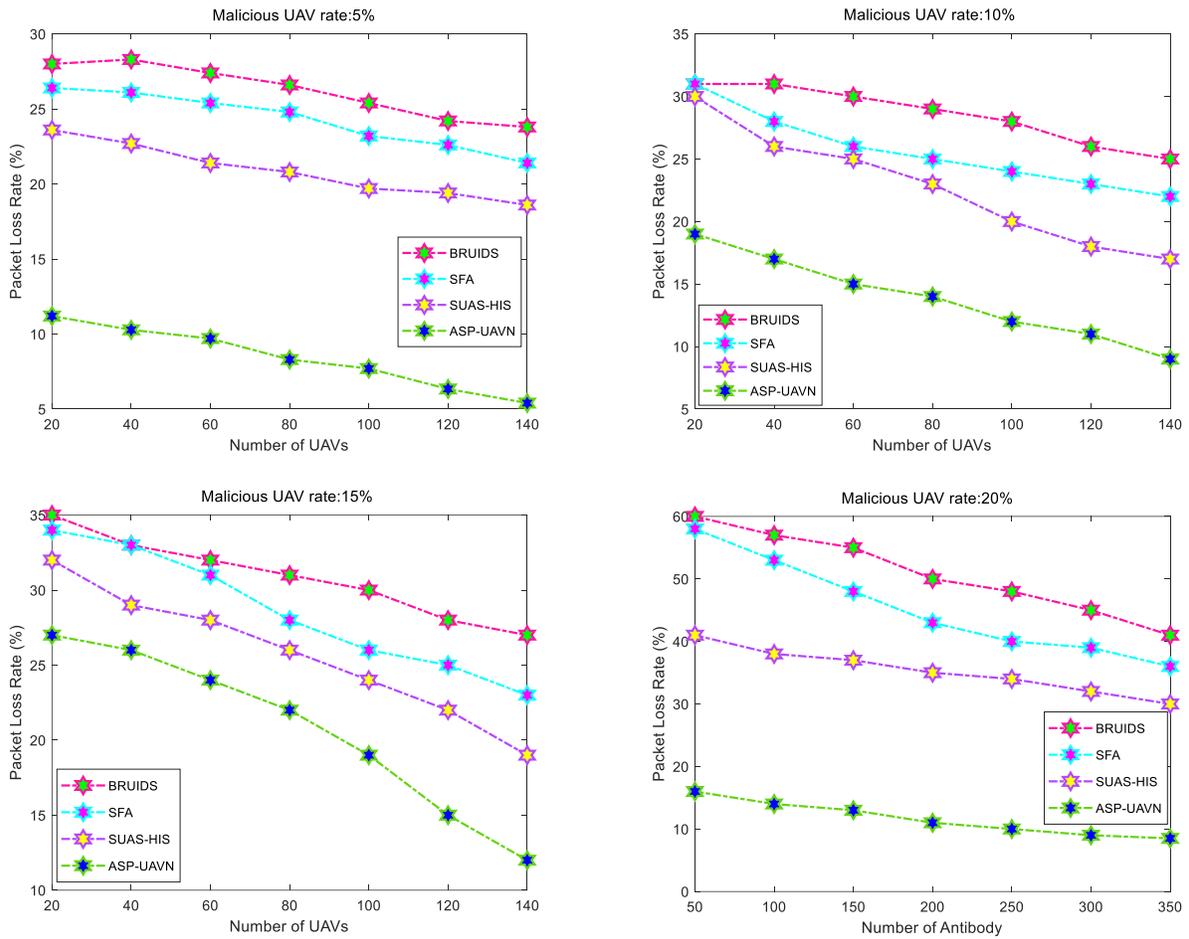

**Fig. 9** Comparison of the ASP-UAVN, SUAS-HIS, SFA and BRUIDS models in term of PLR.

**FP:** Figure 10 presents the false positive rate in four scenarios against normal UAV counts and malicious UAV rates for ASP-UAVN, SUAS-HIS, SFA, and BRUIDS in lethal conditions. As shown in the diagrams, when the number of normal UAVs increases from 20 to 140 and the malicious UAV rate increases from 5 to 20 percent, the generated false positive rate of the proposed method has shown slower and lower growth than other methods. The false positive rate of the proposed method is less than 3 percent when the number of normal UAVs and malicious UAV rate are 120 and 5 percent respectively. However, this value is 17 percent for the SUAS-HIS method, 25 percent for the SFA method, and 35 percent for the BRUIDS method. The reason for the superiority of the proposed scheme is the fast recognition of malicious UAVs and eliminating them with the cooperation of ground stations and normal UAVs using a self-protective method based on AIS. The aforementioned process is carried out using pre-trained rules saved in safety memory.

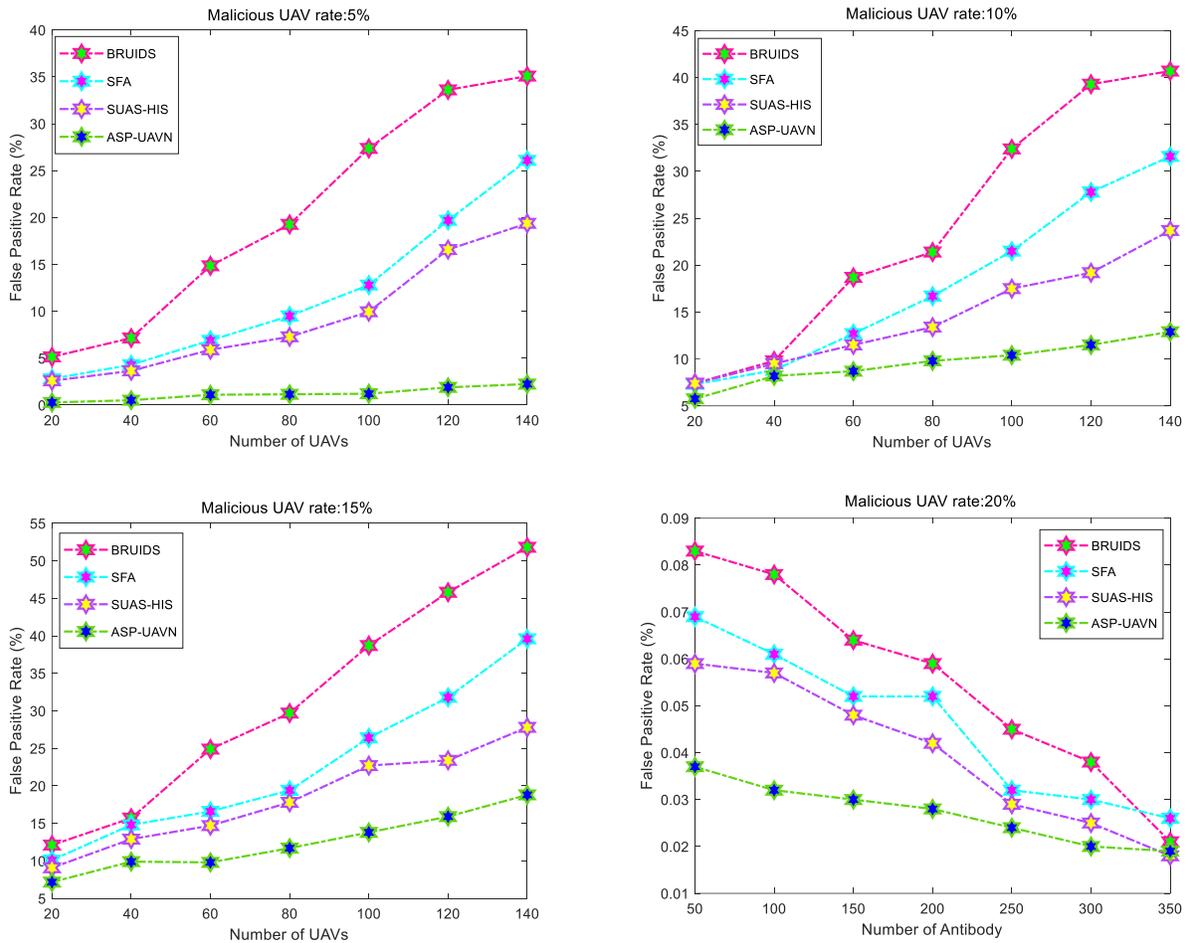

**Fig. 10** Comparison of the ASP-UAVN, SUAS-HIS, SFA and BRUIDS models in term of FP.

**FN:** As shown in the diagrams, the false negative rate (FN) has shown little growth while this value is much higher for ASP-UAVN, SUAS-HIS, SFA, and BRUIDS. In Figure 11, the false negative rate of the proposed method is less than 1.5 percent when the number of UAVs is 120 but for the other three methods, it is 12 percent, 15 percent, and 18 percent respectively. Also, in figure 11, when the number of antibodies is 350, FN is 0.04 in the proposed scheme. However, this value for the other three methods is 0.05, 0.06, and 0.07 respectively. The reason for the FN of the proposed method being low is the utilization of three evaluation, decision-making, and defensive agents that quickly detect malicious UAVs and remove them from the packet transmission operation among UAVs.

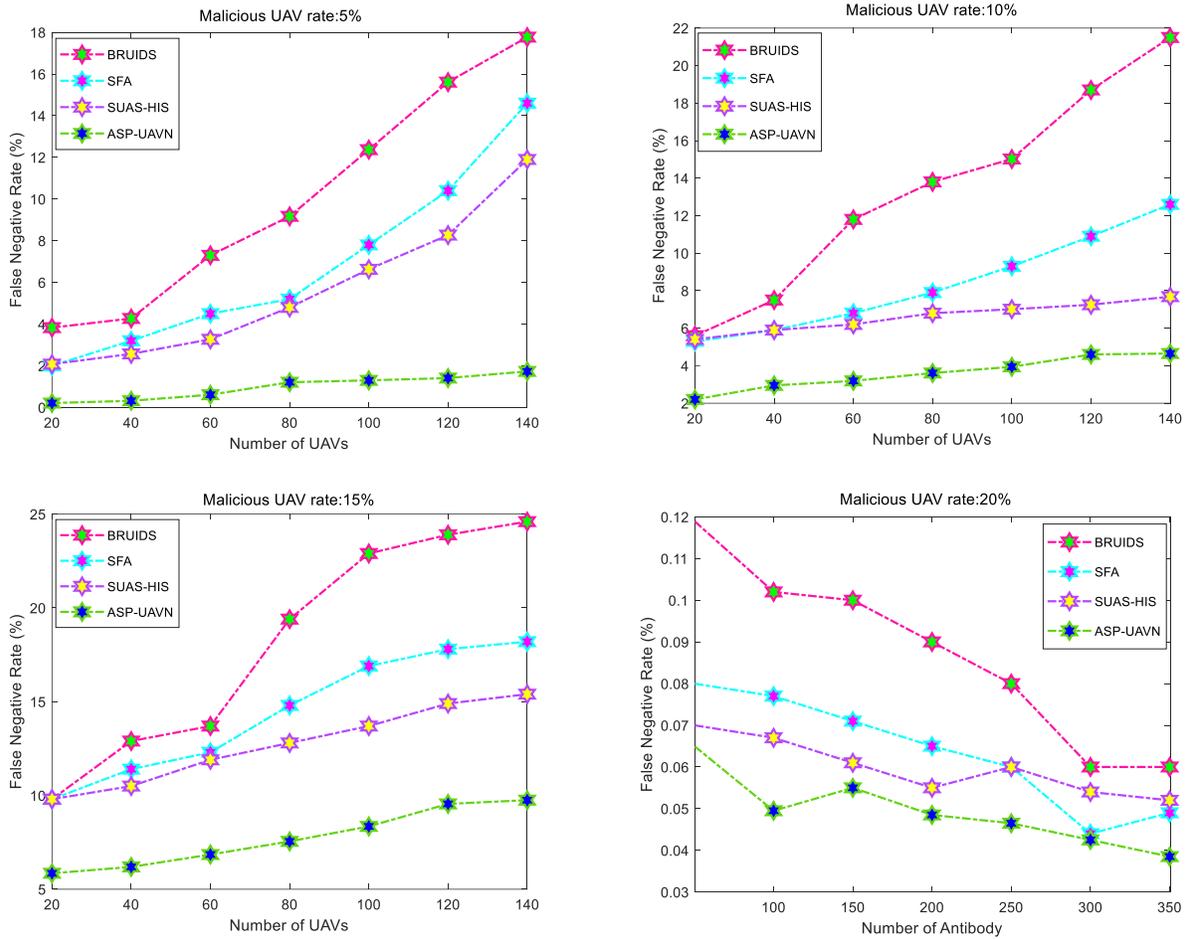

**Fig. 11** Comparison of the ASP-UAVN, SUAS-HIS, SFA and BRUIDS models in term of FN.

**DR:** As shown in the diagrams 12, detection rate (DR) has decreased in all four methods according to the four scenarios, especially when the number of attackers is high. This decrease is much more for BRUIDS compared to other mechanisms. The proposed scheme can detect all the aforementioned attacks with a detection rate higher than 95 percent. This result is achieved when the number of normal UAVs and the malicious UAV rate are 120 and 5 percent respectively. The reason for the superiority of the proposed scheme is the fast identification of malicious UAVs and their elimination using the mapping performed in this scheme. This mapping is carried out between insecure routes defined as anti-genes and the pattern trained based on antibodies. This results in the identification of malicious UAVs and their elimination from the operation cycle.

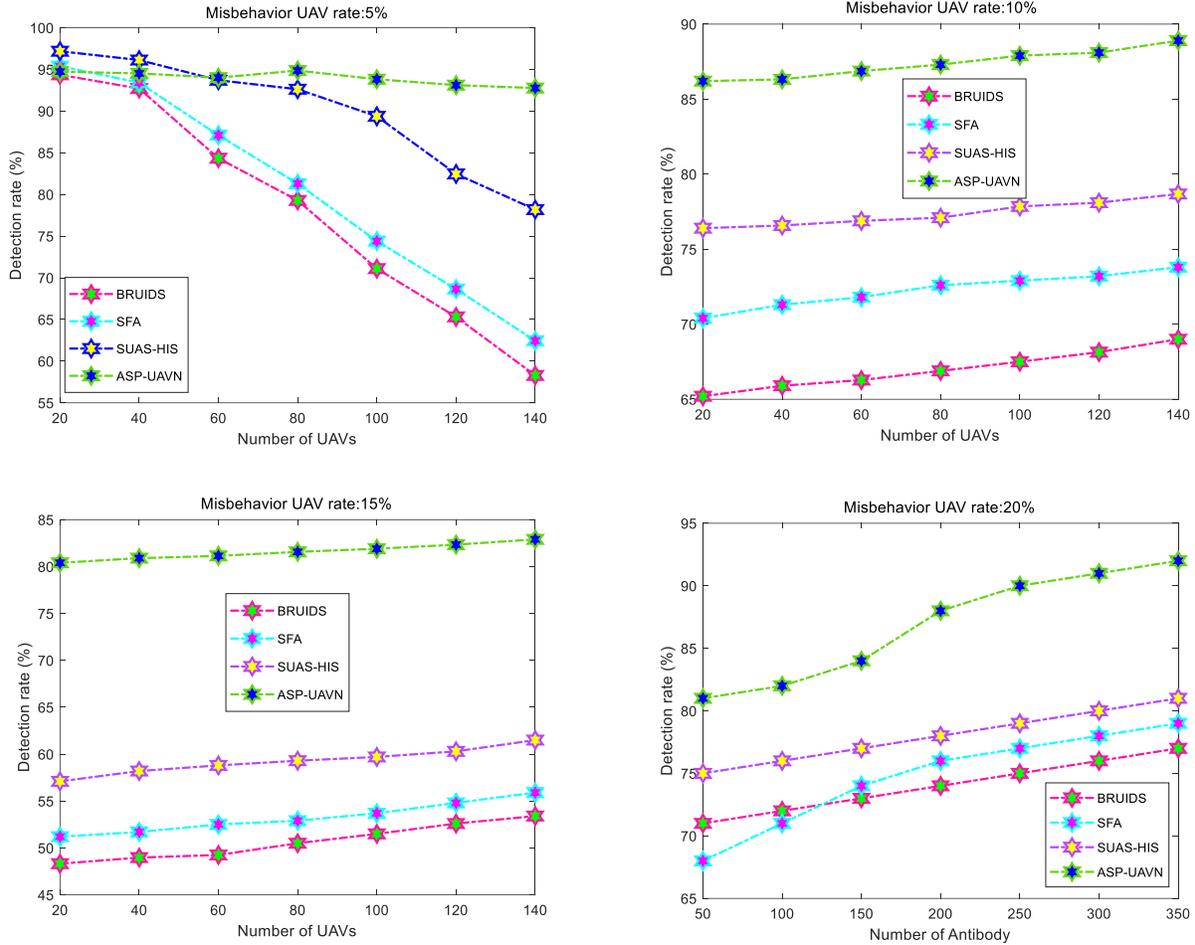

**Fig. 12** Comparison of the ASP-UAVN, SUAS-HIS, SFA and BRUIDS models in term of DR.

# 7 Conclusion and Future work

UAVs have increasing utilization in civilian and military applications recently. Communication security is one of the critical factors ensuring the appropriate UAVs' operation. Rather, UAVs can be captured by adversaries. In this study, it was confirmed that some devastating attacks can be launched simply at a low-cost including wormhole, Sybil and sinkhole attacks, which appears complicated. Hence, it is important to take into account the communication security for UAVs severely. In ASP-UAVN proposed method, the safest route from the source UAV to the destination UAV is chosen according to a self-protective system. In this method, a multi-agent system using an artificial immune system is employed to detect the attacking UAV and choose the safest route. In the proposed P-method, the route request packet (RREQ) is initially transmitted from the source UAV to the destination UAV to detect the existing routes. Then, once the route response packet (RREP) is received, a self-protective method using agents and the knowledge base is employed to choose the safest route and detect the attacking UAVs.

The main advantage of the ASP-UAVN is that the suspect node can be regarded as a normal UAV in the network again followed by a rational penalty. Here, we assessed the ASP-UAVN scheme performance utilizing NS-3 and indicated its high level of detection rate and security (more than 94.5%), low FN (less than 4.95%), low FP (less than 07.104%), and high PDR (over 87.8%), in comparison with the present techniques. In future work, the Firefly algorithm will be used to cluster UAVs and an authentication mechanism will be used to validate UAVs on two security levels to prevent attacks. It is also recommended to use Firefly optimization to reduce power consumption and malicious attacks on drones.

## Conflict of Interest

None.